\documentclass[11pt]{article}
\usepackage{iap2000,epsfig}
\def\gtrsim{\mathrel{\hbox{\rlap{\hbox{\lower4pt\hbox{$\sim$}}}\hbox{$>$}}}}
\let\ga=\gtrsim
\def\lesssim{\mathrel{\hbox{\rlap{\hbox{\lower4pt\hbox{$\sim$}}}\hbox{$<$}}}}
\let\la=\lesssim

\begin{document}
\noindent To appear in {\it Constructing the Universe with Clusters of
Galaxies}, edited by F. Durret \& D. \hspace*{0.3in} Gerbal, in press
\vspace*{3.7cm}
\title{SHOCKS AND NONTHERMAL PROCESSES IN CLUSTERS}

\author{CRAIG L. SARAZIN}

\address{Department of Astronomy, University of Virginia, P. O. Box 3818, \\
 Charlottesville, VA 22903-0818, U.S.A.}

\maketitle\abstracts{
Clusters of galaxies generally form by the gravitational merger of
smaller clusters and groups.
Major cluster mergers are the most energetic events in the Universe
since the Big Bang.
Mergers drive shocks into the intracluster gas, and
these shocks heat the intracluster gas, and should also accelerate
nonthermal relativistic particles.
The X-ray signatures of the thermal effects of merger shocks
will be discussed.
X-ray observations of shocks can be used to determine the geometry and
kinematics of the merger.
As a result of particle acceleration in shocks,
clusters of galaxies should contain very large populations of
relativistic electrons and ions.
Electrons with Lorentz factors $\gamma \sim 300$
(energies $E = \gamma m_e c^2 \sim 150$ MeV) are expected to
be particularly common.
Observations and models for the radio, extreme
ultraviolet, hard X-ray, and gamma-ray emission from nonthermal
particles accelerated in these shocks will also be described.
The predicted gamma-ray fluxes of clusters should make them easily
observable with GLAST.}

\section{Introduction}\label{sec:intro}

Major cluster mergers are the most energetic events in the Universe
since the Big Bang.
Cluster mergers are the mechanism by which clusters are assembled.
In these mergers, the subclusters collide at velocities of
$\sim$2000 km/s, and shocks are driven into the intracluster medium.
In major mergers, these hydrodynamical shocks dissipate energies of
$\sim 3 \times 10^{63}$ ergs; such shocks are the major heating
source for the X-ray emitting intracluster medium.
The shock velocities in merger shocks are similar to those in
supernova remnants in our Galaxy, and we expect them to produce similar
effects.
Mergers shocks should heat and compress the X-ray emitting intracluster
gas, and increase its entropy.
We also expect that particle acceleration by these shocks will produce
nonthermal electrons and ions, and these can produce synchrotron
radio, inverse Compton (IC) EUV and hard X-ray, and gamma-ray emission.

\begin{figure}[t]
\vskip3.8truein
\includegraphics{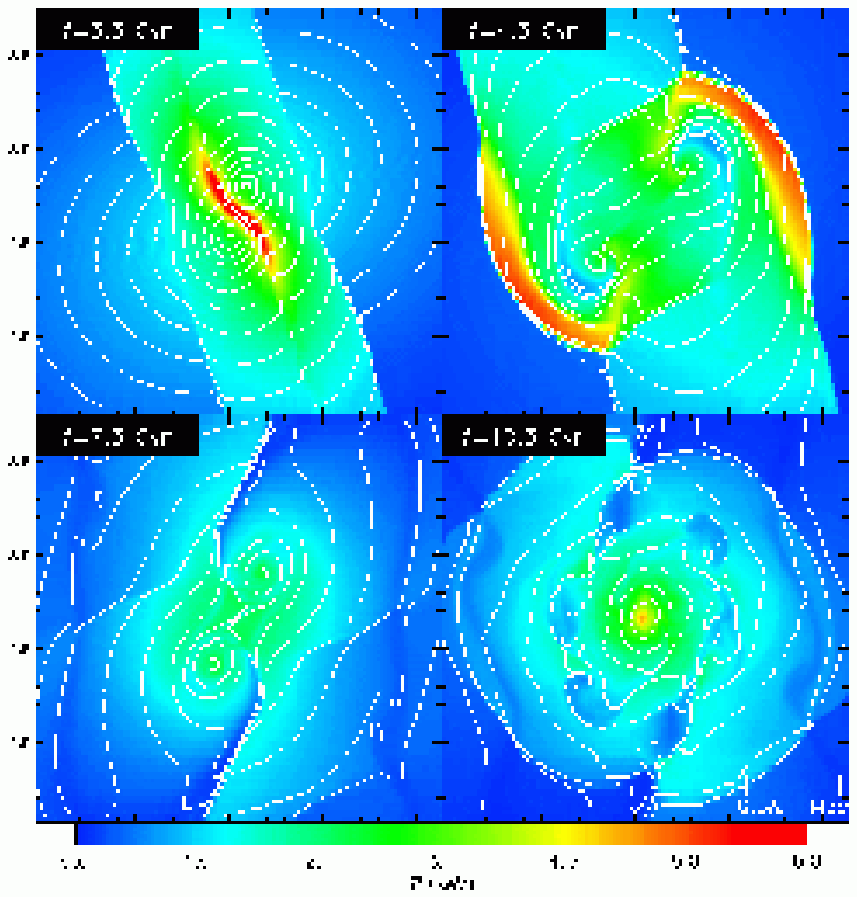}
\caption[]{The results of a hydro simulation of a symmetric, off-center
merger by Ricker and Sarazin.\protect\cite{RS}
The colors show the temperature, while the contours are the X-ray surface
brightness.
Initially, the shocked region is located between the two subcluster
centers.
Later, the main merger shocks propogate to the outer parts of the cluster,
and other weaker shocks also occur.
By the end of the simulation, the cluster is beginning to return to
equilibrium.\hfill
\label{fig:merger_color}}
\end{figure}

Hydrodynamical simulations of cluster formation and evolution have shown
the importance of merger shocks.\cite{SM,RSB}
The evolution of the structure of merger shocks is illustrated in
Figure~\ref{fig:merger_color},\cite{RS}
which shows an off-center merger between two symmetric subclusters.
At early stages in the merger (prior to the first panel), the shocked
region is located between the two subcluster centers and is bounded
on either side by two shocks.
At this time, the subcluster centers, which may contain cooling cores
and central radio sources, are not affected.
Later, these shocks sweep over the subcluster centers (the first panel);
the survival of central cool cores depends on how sharply peaked the
potentials of the subclusters are.\cite{MSV,RS}
The main merger shocks pass into the outer parts of the merging system
(panel 2),
and secondary shocks may appear in the inner regions (panel 3).
Eventually, the cluster begins to return to equilibrium (panel 4).

\section{Thermal Effects of Merger Shocks}\label{sec:thermal}

\subsection{Shock Kinematics}\label{sec:thermal_kinematics}

\begin{figure}[t]
\includegraphics{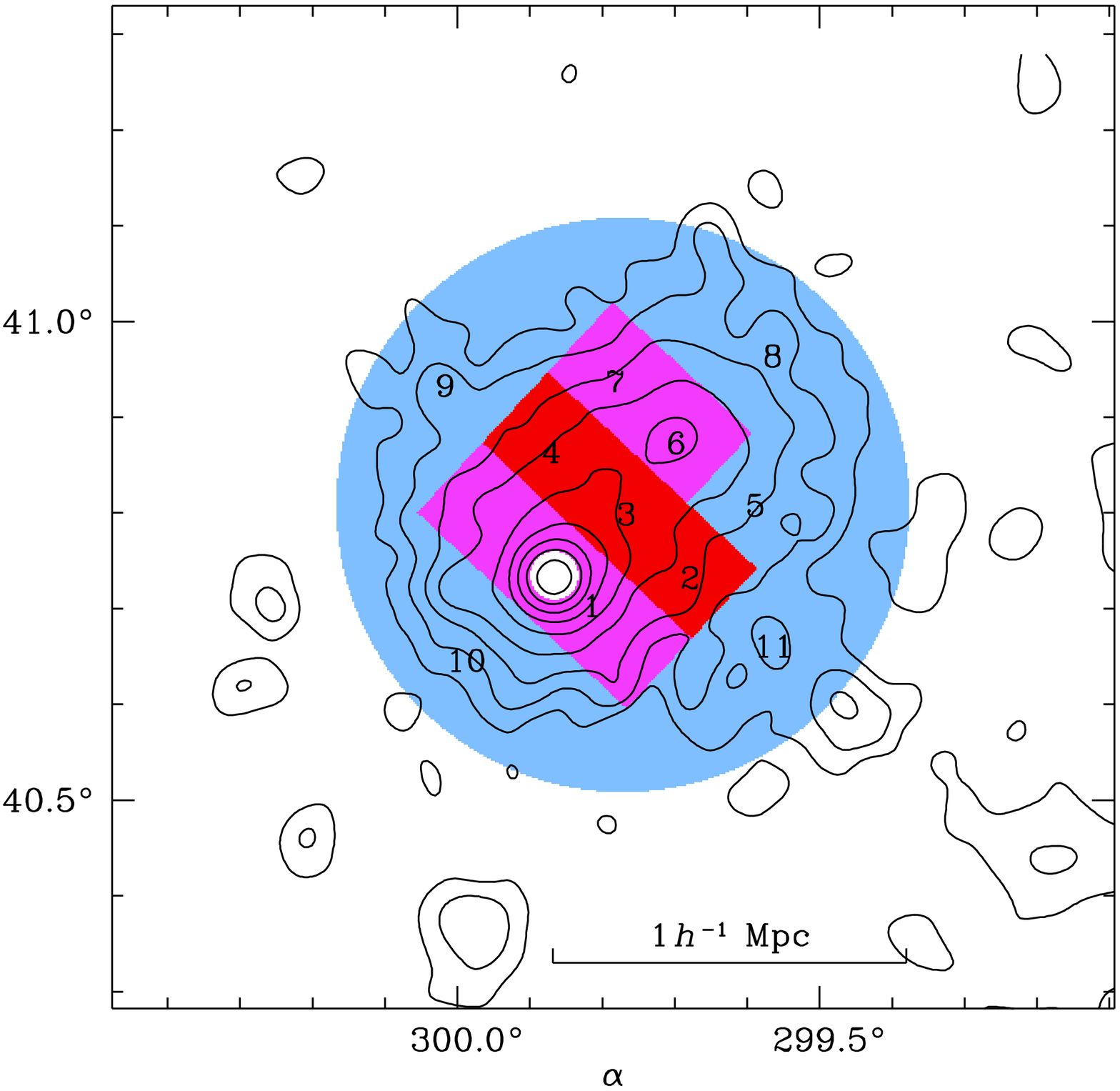}
\includegraphics{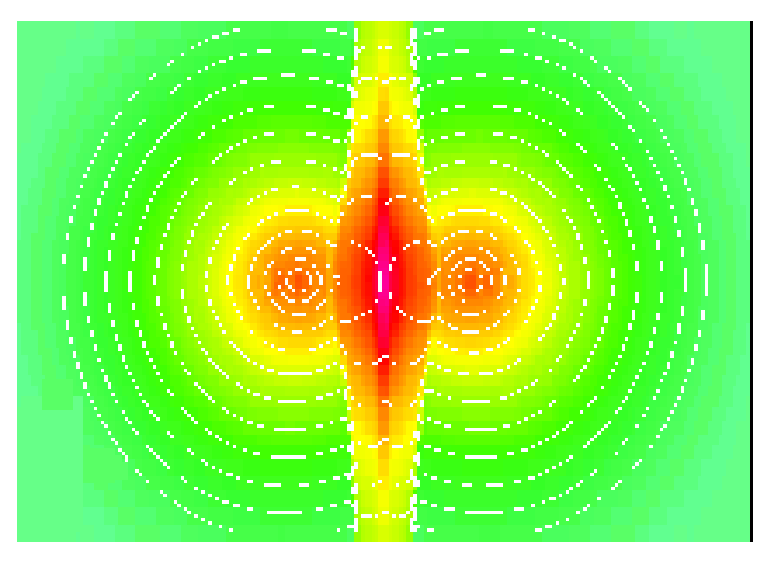}
\centerline{\null}
\centerline{\null}
\noindent\hskip2.44truein (a) \hskip2.75truein (b) \hfill
\vskip2.24in
\caption[]{(a) ROSAT PSPC contours of the Cygnus A cluster.
The radio galaxy is located at the center of the cooling flow peak in
the southeast subcluster.
A second subcluster is centered on the X-ray peak 11$^\prime$ to the
northwest (labeled 6).
The color shows the ASCA temperatures (red being hot, blue being cool),
and indicates that a merger shock region is located between the
two subcluster centers.\cite{MSV}\ 
(b) A hydro/N-body simulation of the Cygnus A cluster merger,\cite{RS}
not to the same scale as panel (a) but rotated to the same orientation.\hfill
\label{fig:cyga}}
\end{figure}

Merger shocks heat and compress the intracluster gas, and these effects
can be used to determine the geometry and kinematics of the merger.
ASCA X-ray temperature maps and ROSAT images have been used in an
initial effort to apply this technique.\cite{MSV}
The cluster containing the radio source Cygnus A is a particularly simple
case (Fig.~\ref{fig:cyga}).
This appears to be a fairly symmetric merger with a low impact parameter.
The merger is at an early phase, with the merger shocks being located
between the two subcluster centers.
A hydro/N-body simulation of the merger\cite{RS} is shown at the right in
Fig.~\ref{fig:cyga} (not to the same scale).
Presumably, the fact that the merger shocks have not yet passed through
the subcluster centers is the reason why the merger hasn't disrupted the
Cygnus A radio source or the surrounding cooling flow.

The simple geometry of this merger makes it easy to apply the shock
jump conditions to determine the merger velocity.
From the Rankine--Hugoniot jump conditions, the velocity change across
the merger shock is 
\begin{equation}
\Delta u_{sh} = \left[\frac{k T_0}{\mu m_p}\left(r-1\right)\left(
\frac{T_1}{T_0}-\frac{1}{r}\right)
\right]^{1/2} \, , \label{eq:shock1}
\end{equation}
where $T_0$ and $T_1$ are the pre- and post-shock temperature,
$\mu$ is the mean mass per particle,
and the shock compression $r$ is
given by
\begin{equation}
\frac{1}{r} =
\left[\frac{1}{4}\left(\frac{\gamma+1}{\gamma-1}\right)^2
\left(\frac{T_1}{T_0}-1\right)^2 +\frac{T_1}{T_0}\right]^{1/2}
-\frac{1}{2}\frac{\gamma+1}{\gamma-1}\left(\frac{T_1}{T_0}-1\right).
\label{eq:shock2}
\end{equation}
For a symmetric merger, the merger velocity of the two subclusters is
just $\Delta u_{cl} = 2 \Delta u_{sh}$.
When the ASCA temperatures are used, this gives
$\Delta u_{cl} \approx 2200$ km/s.
The radial velocity distribution of the galaxies in this cluster
is bimodal,\cite{OLMH} and consistent with a merger velocity of 2400 km/s.

Interestingly, the collision velocity found above is close to the free-fall
velocity of $\sim$2200 km/s that the two subclusters should have achieved
had they fallen from a large distance to their observed separation.
This consistency suggests that the shock energy is effectively thermalized,
and that a major fraction does not go into turbulence, magnetic fields,
or cosmic rays. 

\subsection{Nonequilibrium Effects} \label{sec:thermal_nonequil}

Cluster merger are expected to produce collisionless shocks, as occurs
in supernova remnants.
As such, nonequilibrium effects are expected, including nonequipartition
of electrons and ions and nonequilibrium ionization.\cite{MSV,Tak}
In fact, observations of a central shock in the spectacular merging
cluster Abell~3667 show an apparent lag of $\sim$200 kpc between the
shock position and the peak in the electron temperature.\cite{MSV}
This is consistent with weak electron heating at the shock, and the
timescale for Coulomb heating.
However, recent Chandra observations of this cluster suggest a
more complex situation.\cite{Vea}

\subsection{Early Chandra Results} \label{sec:thermal_chandra}

Although ASCA and ROSAT observations provide some evidence for merger shocks,
they lack the combination of spatial and spectral resolution over a hard
X-ray band needed to really determine shock structures accurately.
Chandra has superb spatial resolution, while XMM has a wider field of
view and a very large collecting area.
Astro-E would have provided very high spectral resolution, sufficient to
directly detect the Doppler motions in the gas.
Hopefully, an equivalent capability will be provided in the not too distant
future.

\begin{figure}[t]
\vskip3.10truein
\includegraphics{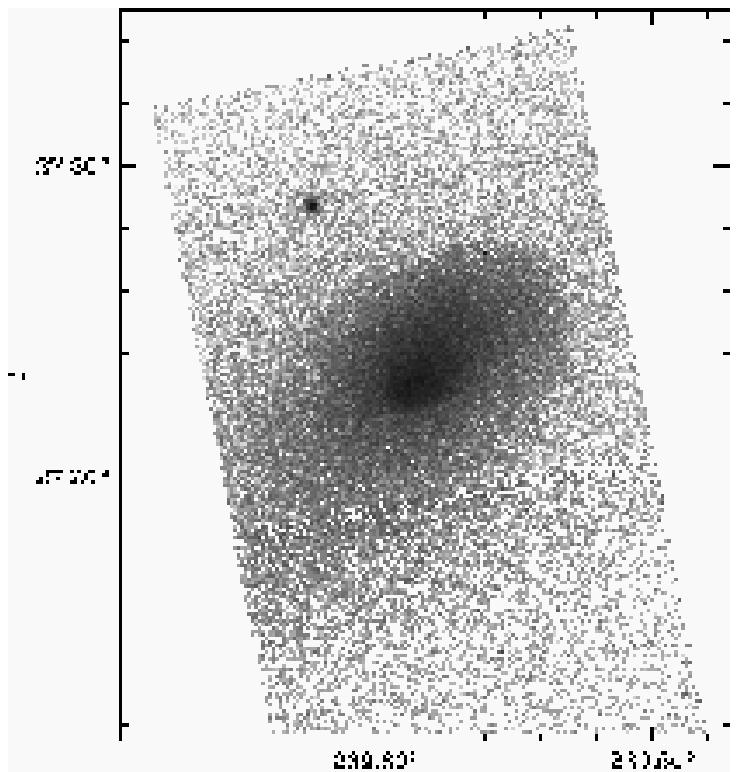}
\caption[]{The Chandra X-ray image of the central region of the merging
cluster Abell~2142 from Markevitch et al.\cite{Mea}\ 
Note the two bow-shaped discontinuities to the northwest and just south
of the brightest region of the cluster.\hfill
\label{fig:a2142}}
\end{figure}

A number of merging clusters were included in the early observations with
Chandra taken for calibration and science verification purposes, and
several of these have shown sharp structures which initially appeared to
be shocks.
An example is Abell~2142 (Fig.~\ref{fig:a2142}), for which a careful
analysis has been given by Markevitch et al.\cite{Mea}
Even the raw Chandra image shows two bow-shaped structures, to the
south and northwest of the center.
However, the Chandra spectra show that these are not shocks.
As one moves inward and the gas density increases abruptly, the gas
temperature decreases in such a way that the pressure is fairly
continuous.
Thus, the specific entropy is actually lower in the denser ``shocked''
region than in the lower density ``pre-shock'' region.
These are not shocks, but rather contact discontinuities between
higher density cool gas and lower density hot gas.
Markevitch et al.\cite{Mea}\ argue that the merger shocks have already
passed though the lower density gas in these regions, and that the
denser cooler gas regions are the remnants of the cooling cores of
the original
subclusters, which are now plowing through the lower density shocked
gas.
A very similar situation has been found in Abell~3667.\cite{Vea}

\section{Particle Acceleration and Nonthermal Emission} \label{sec:nonthermal}

\subsection{Shock Acceleration} \label{sec:nonthermal_acceler}

Radio observations of supernova remnants indicate that
shocks with $v \ga 10^3$ km/s convert at least a few
percent of the shock energy into the acceleration of relativistic
electrons.\cite{BE}
Even more energy may go into relativistic ions.
While the merger shocks in cluster have lower Mach numbers and compressions
than the blast waves in young supernova remnants, it is probable that
shock acceleration operates efficiently in clusters as well.
Given that all of the thermal energy of the intracluster gas in clusters
is due to shocks with such velocities, it seems likely that relativistic
electrons with a total energy of $\ga 10^{62}$ ergs are produced in
clusters, with perhaps even higher energies in ions.

\begin{figure}[t]
\vskip2.20in
\includegraphics{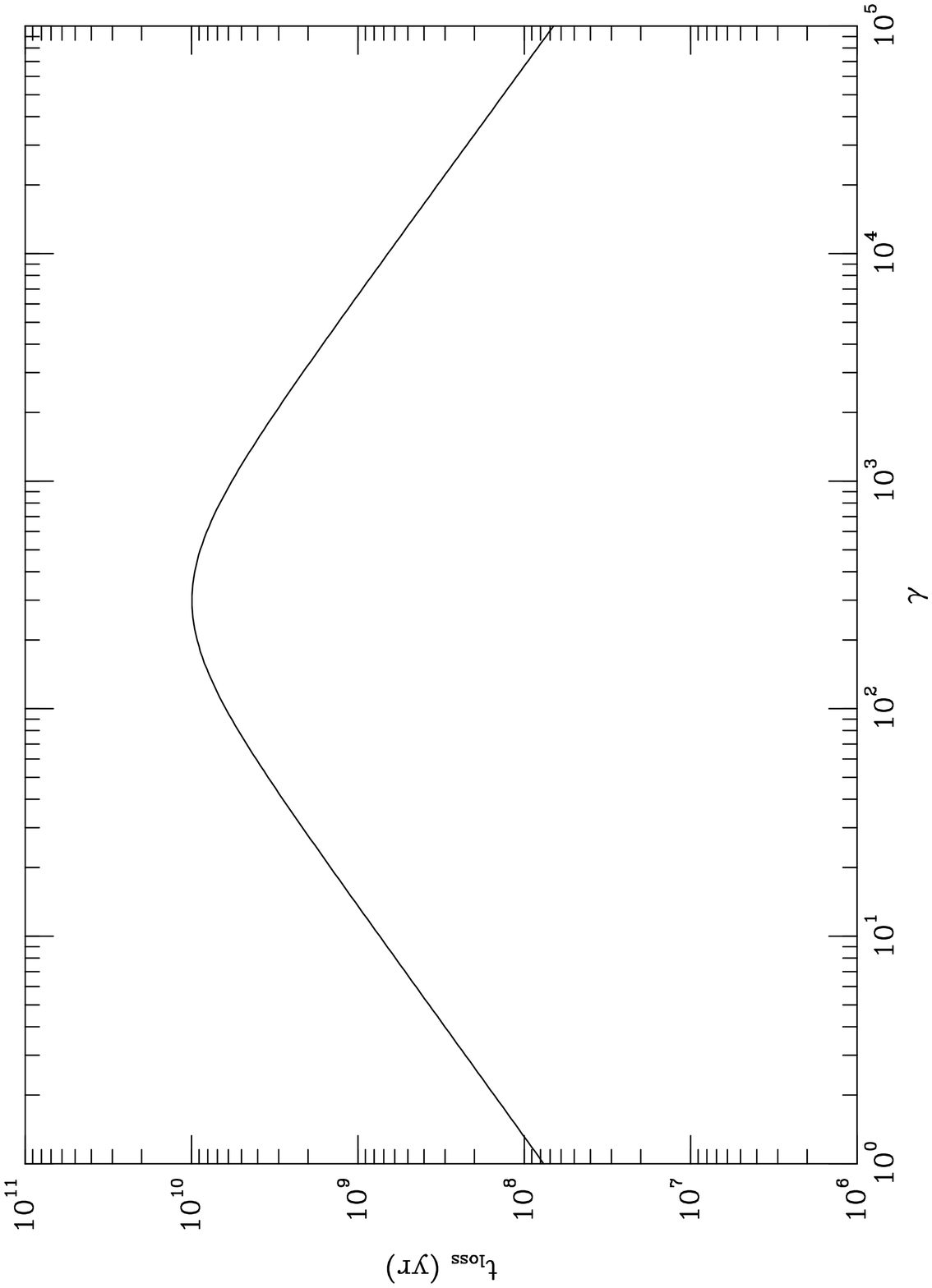}
\caption[]{The loss time scale for relativistic electrons in
a cluster of galaxies under typical intracluster conditions as
a function of their Lorentz factor $\gamma$.
The particles have energies of $E = \gamma m_e c^2$.\hfill
\label{fig:lifetimes}
}
\end{figure}

Clusters are also very good storage locations for cosmic rays.
Under reasonable assumptions for the diffusion coefficient, particles
with energies $\la$$10^6$ GeV have diffusion times which
are longer than the Hubble time.\cite{BBP,CB}
Figure~\ref{fig:lifetimes} gives the loss time scales for relativistic
electrons under typical intracluster conditions.
Although high energy electrons lose energy rapidly due to IC and
synchrotron emission, electrons with Lorentz factors of $\gamma \sim 300$
(energies $\sim 150$ MeV)
have long lifetimes which approach the Hubble time.\cite{S1,SL}
Thus, clusters of galaxies can retain low energy electrons
($\gamma \sim 300$) and nearly all cosmic ray ions for a significant
fraction of a Hubble time.

Recently, I have calculated models for the relativistic electrons in
clusters, assuming they are primary electrons accelerated in merger
shocks.\cite{S1,S2}
An alternative theory is that these particles are secondaries
produced by the decay of charged mesons generated in cosmic ray
ion collisions.\cite{CB}
Two recent cluster merger simulations have included particle acceleration
approximately.\cite{RBS,TN}
Their conclusions are very similar to mine based on simpler models.
The populations of cosmic ray electrons in clusters depends on their merger
histories.
Because low energy electrons have long lifetimes, one expects to find
a large population of them in most clusters (any cluster which has had
a significant merger since $z \sim 1$).
On the other hand, higher energy electrons ($E \ga 1$ GeV) have
short lifetimes (shorter than the time for a merger shock to cross
a cluster).
Thus, one only expects to find large numbers of higher energy electrons
in clusters which are having or have just had a merger.

\begin{figure}[t]
\includegraphics{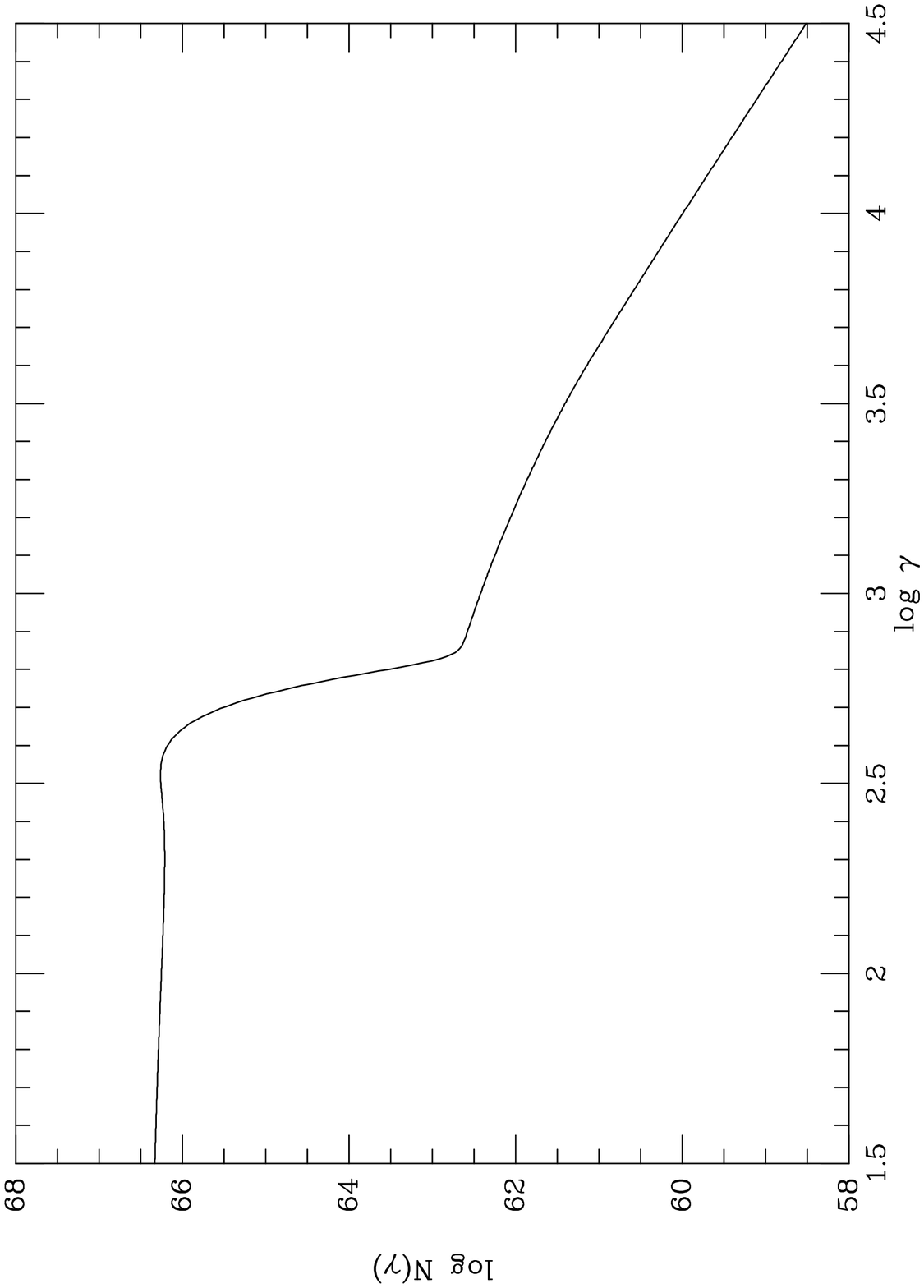}
\includegraphics{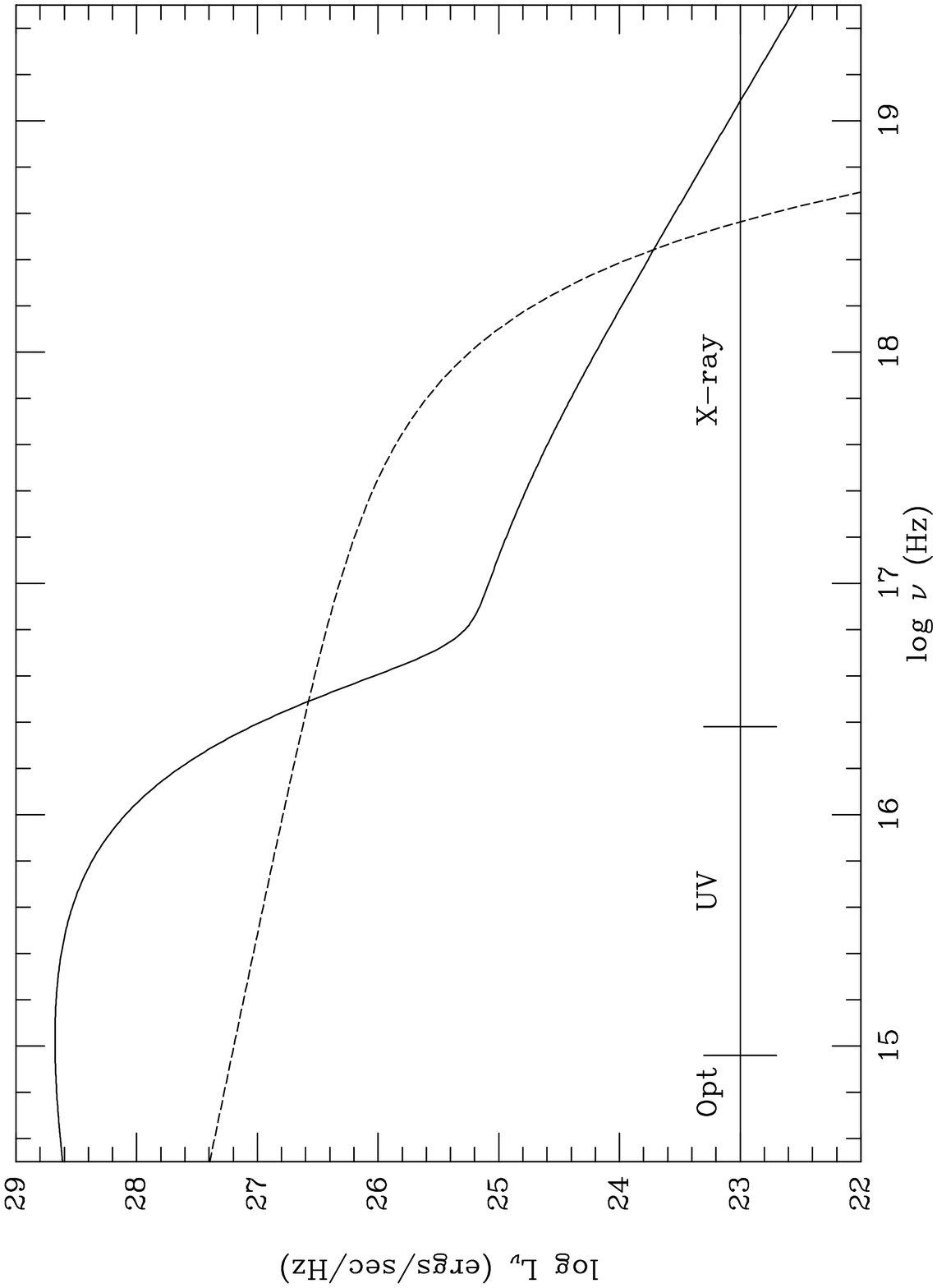}
\centerline{\null}
\centerline{\null}
\noindent\hskip2.66truein (a) \hskip2.94truein (b) \hfill
\vskip1.70in
\caption[]{(a) A typical model for the relativistic electron population
in a cluster of galaxies.
The lower energy electrons are due to all of the mergers in the cluster
history, while the high energy electrons are due to a small current
merger.
(b)
The IC spectrum from the same model (solid curve).
The dashed curve is a 7 keV thermal bremsstrahlung spectrum.\hfill
\label{fig:espect}
}
\end{figure}

Fig.~\ref{fig:espect}(a) shows the electron spectrum in a cluster
with a typical history.
Most of the electron energy is in electrons with $\gamma \sim 300$, which
have the longest lifetimes.
These electrons are produced by mergers over the entire history
of the cluster.
This cluster also has a small ongoing merger which produces
the high energy tail on the electron distribution.

Most of the emission from these electrons is due to IC, and the
resulting spectrum is shown in Fig.~\ref{fig:espect}(b).
For comparison, thermal bremsstrahlung with a typical rich cluster
temperature and luminosity is shown as a dashed curve.
Fig.~\ref{fig:espect}(b) shows that clusters should be strong
sources of extreme ultraviolet (EUV) radiation.
Since this emission is due to electrons with $\gamma \sim 300$ which
have very long lifetimes, EUV radiation should be a common feature
of clusters.\cite{SL}

In clusters with an ongoing merger, the higher energy electrons will
produce a hard X-ray tail via IC scattering of the Cosmic Microwave
Background (CMB);
the same electrons will produce diffuse radio synchrotron emission.

\subsection{EUV/Soft X-ray Emission} \label{sec:nonthermal_euv}

Excess EUV emission has apparently been detected with the EUVE satellite in
six clusters (Virgo, Coma, Abell 1795, Abell 2199, Abell 4038, \&
Abell 4059).\cite{Lea1,Lea2,BB,MLL,BBK2,Kea,Lea3,Lea4,BBK1,BLM,B}
In fact, the EUVE satellite appears to have detected all of the clusters
it observed which are nearby, which have long integration times, and 
which lie in directions of low Galactic column where detection is
possible at these energies.
However, the EUV detections and claimed properties of the clusters
remain quite controversial.\cite{BB,AB,BBK2,BBK1}

\begin{figure}[t]
\vskip2.20in
\includegraphics{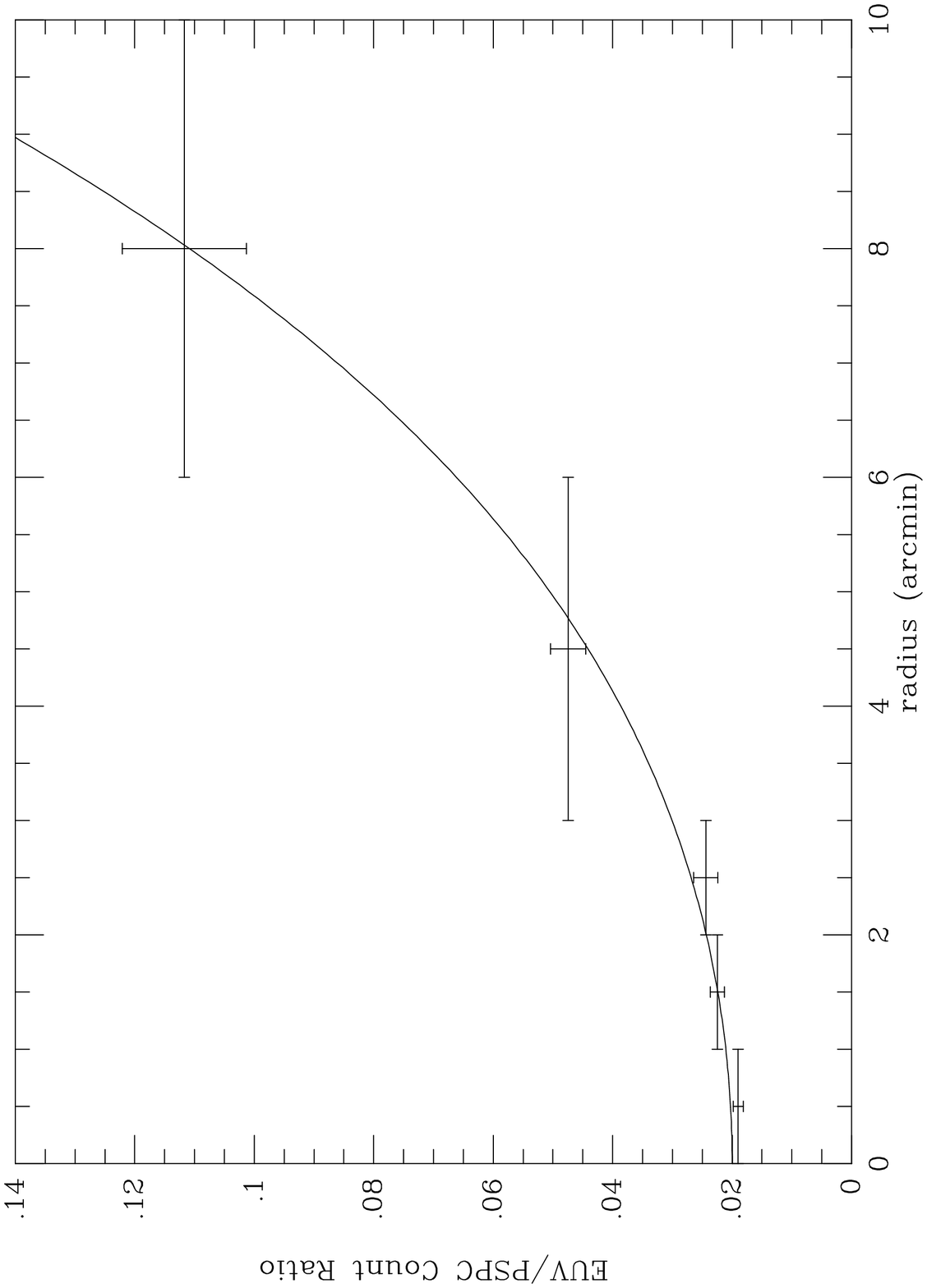}
\caption[]{The data points give the observed ratio of the EUV to
thermal X-ray surface brightness versus radius in the cluster Abell~1795
from Mittaz et al.\cite{MLL}
The EUV data is from EUVE, while the X-ray data is from the ROSAT
PSPC.
The solid curve is the IC model from Sarazin \& Lieu,\cite{SL}
assuming that the pressure in cosmic ray electrons in this cluster varies
in proportion with the thermal pressure in the X-ray emitting gas, as
determined with the ROSAT PSPC.\hfill
\label{fig:euv_distr}}
\end{figure}

The EUV observations suggest that rich clusters generally have
EUV luminosities of $\sim$$10^{44}$ ergs/s, and have spectra which
decline rapidly in going from the EUV to the X-ray band.
The EUV emission is very spatially extended.
The last point is illustrated in Figure~\ref{fig:euv_distr}.
The data points show the ratio of the EUV to thermal X-ray emission
in the cluster Abell~1795.
The ratio increases rapidly with increasing radius.

While it is possible that the EUV emission may be thermal in
origin,\cite{F97,BLM}
I believe that it is more likely that this emission is due to
inverse Compton scattering (IC) of CMB photons by low energy
relativistic electrons.\cite{H97,BB,EB,SL}
In this model, the EUV would be produced by electrons with
energies of $\sim$150 MeV ($\gamma \sim 300$; Fig.~\ref{fig:espect}).
As noted above, these electrons have lifetimes which are comparable to
the Hubble time, and should be present in essentially all clusters.
This can explain why EUV emission is common.
To produce the EUV luminosities observed, one needs a population of
such electrons with a total energy of $\sim$$10^{62}$ ergs, which is
about 3\% of the typical thermal energy content of clusters.
This is a reasonable acceleration efficiency for these particles, given
that both the thermal energy in the intracluster gas and the relativistic
particles result from merger shocks.
The steep spectrum in going from EUV to X-ray bands is predicted by
this model
[Fig.~\ref{fig:espect}(b)],
it results from the rapid increase in losses ($\propto \gamma^2$)
for particles as the energy increases above $\gamma \sim 300$
(Fig.~\ref{fig:lifetimes}).

The broad spatial distribution of the EUV is also naturally explained by the
density dependence of IC emission.
The thermal emission which produces the bulk of the X-ray emission in
clusters is due to collisions between thermal electrons and ions;
thus, it declines with the square of the density as the radius increases.
On the other hand, IC emission is due to collisions between cosmic ray
electrons and CMB photons;
since the CMB energy density is extremely uniform, IC EUV emission
varies with a single power of density, rather than density squared.
This simple difference can explain why the EUV is more extended than
the thermal X-ray emission.
As a example, the solid curve in Figure~\ref{fig:euv_distr} shows the
predicted ratio of EUV to X-ray emission in Abell~1795, if the pressure
in cosmic ray electrons varies in proportion with the thermal gas pressure,
and the latter is determined from the X-ray observations.\cite{SL}

\subsection{Hard X-ray Tails } \label{sec:nonthermal_hxr}

If clusters contain higher energy relativistic electrons with
$\gamma \sim 10^4$, these particles will produce hard X-ray emission
by IC scattering, and radio synchrotron emission depending on the
intracluster magnetic field.
Since these higher energy electrons have short lifetimes, they should
only be present in clusters with evidence for a recent or ongoing merger.

Hard X-ray emission in excess of the thermal emission and detected as
a nonthermal tail at energies $\ga$20 keV has been seen in at least
two clusters.
The Coma cluster, which is undergoing at least one merger and which
has a radio halo, was detected with both BeppoSAX and
RXTE.\cite{FFea1,RGB}
BeppoSAX also has detected Abell~2256,\cite{FFea2}
another merger cluster with strong diffuse radio emission.
BeppoSAX may have detected Abell~2199,\cite{Kea}
although I believe the evidence is less compelling for this case.
A nonthermal hard X-ray detection of Abell~2199 would be surprising,
as this cluster is very relaxed and has no radio halo or relic.\cite{KS1}
An alternative explanation of the hard X-ray tails is that they might be
due to nonthermal bremsstrahlung,\cite{SK}
which is bremsstrahlung from nonthermal electrons with energies of
10--1000 keV which are being accelerated to higher energies.

\subsection{Radio Halos and Relics } \label{sec:nonthermal_radio}

A number of clusters of galaxies are known to contain large-scale
diffuse radio sources which have no obvious connection to individual
galaxies in the cluster.\cite{Gea}
These sources are referred to as radio halos when they appear projected
on the center of the cluster, and are called relics when they are found
on the cluster periphery.
The best known and studied radio halo is in the Coma cluster.\cite{Dea}
These radio halos and relics are relatively uncommon; about 40 are known
at the present time.
They are diffuse radio sources with very steeply declining radio spectra.
In all cases of which I am aware, they have been found in clusters which
show significant evidence for an ungoing merger.\cite{Gea,Fer1,Fer2}
This suggests that the relativistic electrons are accelerated by merger
shocks.
In the early stages of mergers, halos are often found on the border
between the subclusters, where the cluster gas is first being shocked
(e.g., Abell 85).\cite{SR}
In more advanced mergers, more conventional centrally located halos
(e.g., Coma) and peripheral relics (e.g., Abell 3667) are found.
Abell~3667 provides a spectacular case of two bow-shaped radio relics
at opposite sides of a merging cluster,\cite{Rea}
and located at the positions where merger shocks are predicted.\cite{RBS}

Although radio halos and relics have been studied for some time, lately the
number of known cases has increased dramatically.
This is largely due to the radio surveys (like NVSS and WENSS) and
new low frequency radio instruments.
Perhaps the largest number of new sources have come from the NVSS
survey and VLA pointed observation follow up effort by Giovannini,
Feretti, and collaborators.\cite{Gea}
Other new detections have involved other VLA surveys,\cite{OMV}
the VLA at 74 MHz,\cite{KPE}
SUMSS and MOST,\cite{Rea}
the ATCA,\cite{LHBA} and
WENSS.\cite{KS2}

\subsection{Predicted Gamma-Ray Emission} \label{sec:nonthermal_gamma}

Relativistic electrons and ions in clusters are also expected to produce
strong gamma-ray emission.\cite{CB,BC,Bla,S2}
The region near 100 MeV is particularly
interesting, as this region includes bremsstrahlung from the most common
electrons with $\gamma \sim 300$, and gamma-rays from ions produced by
$\pi^o$ decay.
Both of these processes involve collisions between relativistic particles
(electrons for bremsstrahlung, ions for $\pi^o$ emission)
and thermal particles, so they should both vary in the same way with
density in the cluster.
Thus, the ratio of these two spectrally distinguishable emission processes
should tell us the ratio of cosmic ray ions to electrons in
clusters.\cite{Bla,S2}

\begin{figure}[t]
\vskip2.20in
\includegraphics{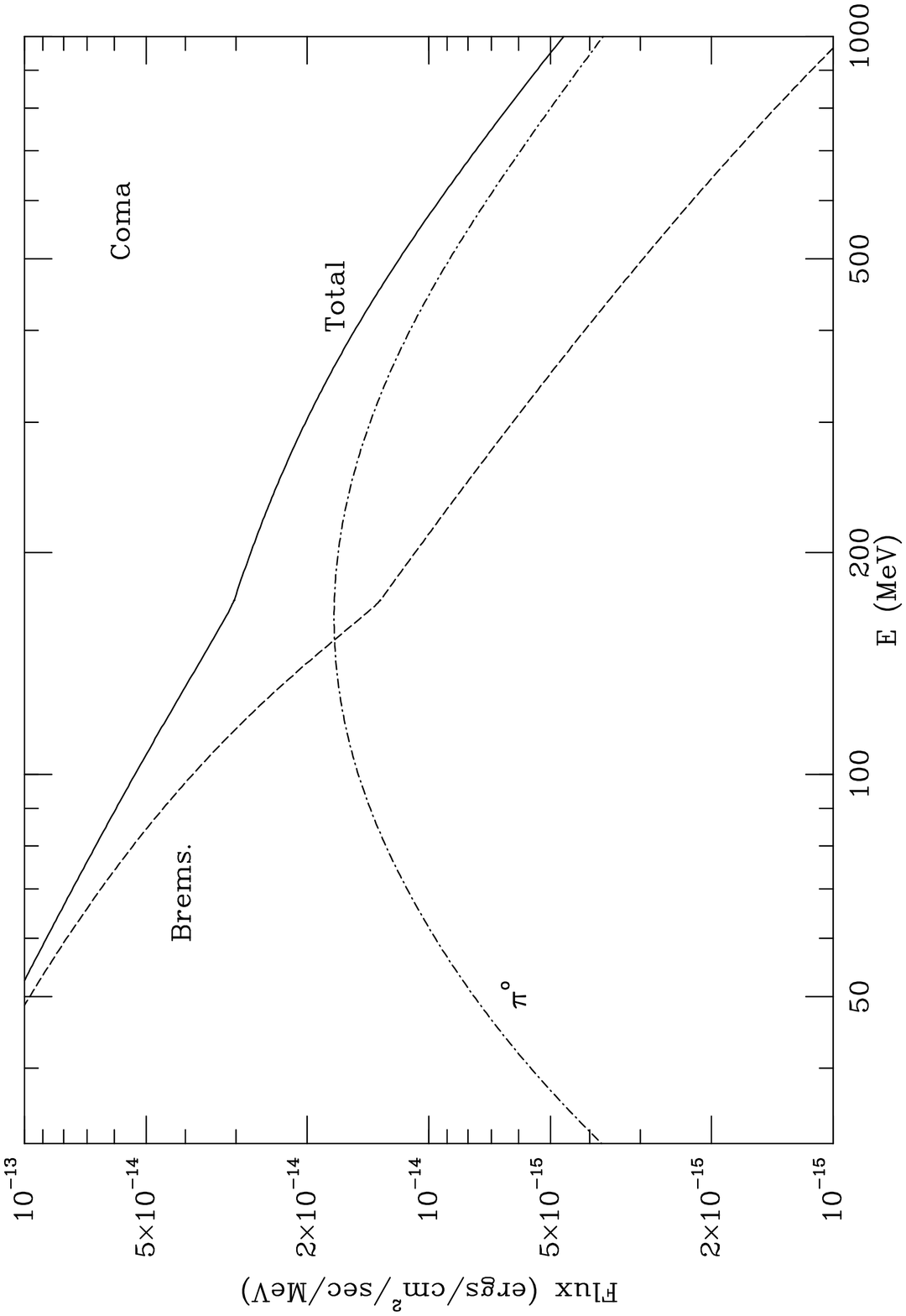}
\caption[]{The predicted gamma-ray spectrum for the Coma cluster, including
electron bremsstrahlung and $\pi^o$ decay from ions.\cite{S2}\hfill
\label{fig:gamma}}
\end{figure}

Figure~\ref{fig:gamma} shows the predicted gamma-ray spectrum for the
Coma cluster, based on a model which reproduces the observed EUV, hard
X-ray, and radio emission.\cite{S2}
The observed upper limit from CGO/EGRET is $<$$4 \times 10^{-8}$
cts/cm$^2$/s for $E > 100$ MeV,\cite{Sea}
while the predicted value for this model is
$\sim$$2 \times 10^{-8}$ cts/cm$^2$/s.
The EGRET upper limit already shows that the ratio of ions to electrons
cannot be too large ($\la$30).\cite{Bla,S2})
The predicted fluxes are such that many nearby clusters should be easily
detectable with GLAST.

\section*{Acknowledgments}
I want to thank my collaborators Maxim Markevitch, Paul Ricker,
Alexey Vikhlinin, and Josh Kempner for all their help.
Maxim Markevitch very kindly provided his Chandra image of Abell~2142
(Fig.~\ref{fig:a2142}) for this paper and my talk,
and Alexey Vikhlinin provided his image of Abell~3667 for my talk.
Paul Ricker and Josh Kempner helped with several of the figures.
Support for this work was provided by the National Aeronautics and Space
Administration through Chandra Award Numbers GO0-1019X and GO0-1141X issued
by the Chandra X-ray Observatory Center, which is operated by the Smithsonian
Astrophysical Observatory for and on behalf of NASA under contract
NAS8-39073.


\section*{References}
\begin{thebibliography}{99}


\bibitem{AB} Arabadjis, J. S., \& Bregman, J. N. 1999, ApJ, 514, 607

\bibitem{BBP} Berezinsky, V. S., Blasi, P., \& Ptuskin, V. S. 1997, ApJ, 487, 529

\bibitem{BBK1} Bergh\"ofer, T. W., Bowyer, S., \& Korpela, E. 2000, ApJ, 535, 615

\bibitem{Bla} Blasi, P. 1999, ApJ, 525, 603

\bibitem{BC} Blasi, P., \& Colafrancesco, S. 1999, APh, 12, 169

\bibitem{BLM} Bonamente, M., Lieu, R., \& Mittaz, J. 2000, ApJ, in press

\bibitem{B} Bowyer, S. 2000, private communication

\bibitem{BB} Bowyer, S., \& Bergh\"ofer, T. W. 1998, ApJ, 506, 502

\bibitem{BBK2} Bowyer, S., Bergh\"ofer, T. W., \& Korpela, E. 1999, ApJ, 526, 592

\bibitem{BE} Blandford, R. D., \& Eichler, D. 1987, Phys.\ Rep., 154, 1

\bibitem{CB} Colafrancesco, S., \& Blasi, P. 1998, APh, 9, 227

\bibitem{Dea} Deiss, B. M., Reich, W., Lesch, H., \& Wielebinski, R.
1997, A\&A, 321, 55D

\bibitem{EB} En{\ss}lin, T. A., \& Biermann, P. L. 1998, A\&A, 330, 90

\bibitem{F97} Fabian, A. C. 1997, Science, 275, 48

\bibitem{Fer1} Feretti, L. 1999, in Proc.\ Diffuse Thermal and Relativistic
Plasma in Galaxy Clusters, ed.\ H. B\"{o}ringer, L. Feretti, \& P. Schuecker
(Garching: MPE), 1

\bibitem{Fer2} Feretti, L. 2000, preprint (astro-ph/0006379)

\bibitem{FFea1} Fusco-Femiano, R., et al., 1999, ApJ, 513, L21

\bibitem{FFea2} Fusco-Femiano, R., et al., 2000, ApJ, 534, L7

\bibitem{Gea} Giovannini, G., et al., 1993, ApJ, 406, 399

\bibitem{GTF} Giovannini, G., Tordi, M., \& Feretti, L. 1999, New Astronomy, 4, 14

\bibitem{H97} Hwang, C.-Y. 1997, Science, 278, 1917

\bibitem{Kea} Kaastra, J. S., et al., 1999, ApJ, 519, L119

\bibitem{KPE} Kassim, N. E., Perley, R. A., \& Erickson, W. C. 1999, in
Proc.\ Diffuse Thermal and Relativistic
Plasma in Galaxy Clusters, ed.\ H. B\"{o}ringer, L. Feretti, \& P. Schuecker
(Garching: MPE), 49

\bibitem{KS1} Kempner, J., \& Sarazin, C. L. 2000a, ApJ, 530, 282

\bibitem{KS2} Kempner, J., \& Sarazin, C. L. 2000b, ApJ, submitted

\bibitem{LHBA} Liang, H., Hunstead, R. W., Birkinshaw, M., \& Andreani, P.
2000, ApJ, in press

\bibitem{Lea1} Lieu, R., et al., 1996, Science, 274, 1335

\bibitem{Lea2} Lieu, R., et al., 1996, ApJ, 458, L5

\bibitem{Lea3} Lieu, R., Bonamente, M., \& Mittaz, J. 1999, ApJ, 517, L91

\bibitem{Lea4} Lieu, R., et al., 1999, ApJ, 527, L77

\bibitem{MSV} Markevitch, M., Sarazin, C. L., \& Vikhlinin, A. 1999, ApJ, 521, 526

\bibitem{Mea} Markevitch, M., et al., 2000, ApJ, in press

\bibitem{MLL} Mittaz, J. P. D., Lieu, R., \& Lockman, F. J. 1998, ApJ, 498, L17

\bibitem{OLMH} Owen, F. N., et al., 1997, ApJ, 488, L15

\bibitem{OMV} Owen, F. N., Morrison, G., \& Voges, W. in
Proc.\ Diffuse Thermal and Relativistic
Plasma in Galaxy Clusters, ed.\ H. B\"{o}ringer, L. Feretti, \& P. Schuecker
(Garching: MPE), 9

\bibitem{RGB} Rephaeli, Y., Gruber, D., \& Blanco, P. 1999, ApJ, 511, L21

\bibitem{RS} Ricker, P. M., \& Sarazin, C. L. 2000, preprint

\bibitem{RBS} Roettiger, K., Burns, J., \& Stone, J. M. 1999, ApJ, 518, 603

\bibitem{RSB} Roettiger, K., Stone, J. M., \& Burns, J. 1999, ApJ, 518, 594

\bibitem{Rea} R\"ottgering, H., Wieringa, M., Hunstead, R., \& Ekers, R. 1997, MNRAS, 290, 57

\bibitem{S1} Sarazin, C. L. 1999, ApJ, 520, 529

\bibitem{S2} Sarazin, C. L. 2000, preprint

\bibitem{SK} Sarazin, C. L., \& Kempner, J. 2000, ApJ, in press

\bibitem{SL} Sarazin, C. L., \& Lieu, R. 1998, ApJ, 494, L177

\bibitem{SM} Schindler, S., \& M\"uller E. 1993, A\&A, 272, 137

\bibitem{SR} Slee, O. B., \& Reynolds, J. E. 1984, PASA, 5, 516

\bibitem{Sea} Sreekumar, P., et al., 1996, ApJ, 464, 628

\bibitem{Tak} Takizawa, M. 1999, ApJ, 520, 514

\bibitem{TN} Takizawa, M., \& Naito, T. 2000, ApJ, 535, 586

\bibitem{Vea} Vikhlinin, A., et al., 2000, preprint

\end{thebibliography}
\end{document}